\documentstyle[twoside,graphicx,alltt]{article}

\catcode`\@=11
\long\def\@makefntext#1{
\protect\noindent \hbox to 3.2pt {\hskip-.9pt  
$^{{\eightrm\@thefnmark}}$\hfil}#1\hfill}               

\def\@makefnmark{\hbox to 0pt{$^{\@thefnmark}$\hss}}    
        
\def\ps@myheadings{\let\@mkboth\@gobbletwo
\def\@oddhead{\hbox{}
\rightmark\hfil\eightrm\thepage}   
\def\@oddfoot{}\def\@evenhead{\eightrm\thepage\hfil
\leftmark\hbox{}}\def\@evenfoot{}
\def\sectionmark##1{}\def\subsectionmark##1{}}

\oddsidemargin=\evensidemargin
\newcounter{sectionc}\newcounter{subsectionc}\newcounter{subsubsectionc}
\renewcommand{\section}[1] {\vspace{12pt}\addtocounter{sectionc}{1} 
\setcounter{subsectionc}{0}\setcounter{subsubsectionc}{0}\noindent 
        {\tenbf\thesectionc. #1}\par\vspace{5pt}}
\renewcommand{\subsection}[1] {\vspace{12pt}\addtocounter{subsectionc}{1} 
        \setcounter{subsubsectionc}{0}\noindent 
        {\bf\thesectionc.\thesubsectionc. {\kern1pt \bfit #1}}\par\vspace{5pt}}
\renewcommand{\subsubsection}[1] {\vspace{12pt}\addtocounter{subsubsectionc}{1}
        \noindent{\tenrm\thesectionc.\thesubsectionc.\thesubsubsectionc.
        {\kern1pt \tenit #1}}\par\vspace{5pt}}
\newcommand{\nonumsection}[1] {\vspace{12pt}\noindent{\tenbf #1}
        \par\vspace{5pt}}

\newcounter{appendixc}
\newcounter{subappendixc}[appendixc]
\newcounter{subsubappendixc}[subappendixc]
\renewcommand{\thesubappendixc}{\Alph{appendixc}.\arabic{subappendixc}}
\renewcommand{\thesubsubappendixc}
        {\Alph{appendixc}.\arabic{subappendixc}.\arabic{subsubappendixc}}

\renewcommand{\appendix}[1] {\vspace{12pt}
        \refstepcounter{appendixc}
        \setcounter{figure}{0}
        \setcounter{table}{0}
        \setcounter{lemma}{0}
        \setcounter{theorem}{0}
        \setcounter{corollary}{0}
        \setcounter{definition}{0}
        \setcounter{equation}{0}
        \renewcommand{\thefigure}{\Alph{appendixc}.\arabic{figure}}
        \renewcommand{\thetable}{\Alph{appendixc}.\arabic{table}}
        \renewcommand{\theappendixc}{\Alph{appendixc}}
        \renewcommand{\thelemma}{\Alph{appendixc}.\arabic{lemma}}
        \renewcommand{\thetheorem}{\Alph{appendixc}.\arabic{theorem}}
        \renewcommand{\thedefinition}{\Alph{appendixc}.\arabic{definition}}
        \renewcommand{\thecorollary}{\Alph{appendixc}.\arabic{corollary}}
        \noindent{\tenbf Appendix \theappendixc #1}\par\vspace{5pt}}
\newcommand{\subappendix}[1] {\vspace{12pt}
        \refstepcounter{subappendixc}
        \noindent{\bf Appendix \thesubappendixc. {\kern1pt \bfit #1}}
        \par\vspace{5pt}}
\newcommand{\subsubappendix}[1] {\vspace{12pt}
        \refstepcounter{subsubappendixc}
        \noindent{\rm Appendix \thesubsubappendixc. {\kern1pt \tenit #1}}
        \par\vspace{5pt}}

\newcommand{\textlineskip}{\baselineskip=13pt}
\newcommand{\smalllineskip}{\baselineskip=10pt}

\def\eightcirc{
\begin{picture}(0,0)
\put(4.4,1.8){\circle{6.5}}
\end{picture}}
\def\eightcopyright{\eightcirc\kern2.7pt\hbox{\eightrm c}} 

\newcommand{\copyrightheading}[1]
        {\vspace*{-2.5cm}\smalllineskip{\flushleft
        {\footnotesize International Journal of Modern Physics C #1}\\
        {\footnotesize $\eightcopyright$\, World Scientific Publishing
         Company}\\
         }}

\newcommand{\publisher}[2]{{\begin{center}\footnotesize\smalllineskip 
        Received #1\\
        Revised #2
        \end{center}
        }}

\def\abstracts#1#2#3{{
        \centering{\begin{minipage}{4.5in}\footnotesize\baselineskip=10pt
        \parindent=0pt #1\par 
        \parindent=15pt #2\par
        \parindent=15pt #3
        \end{minipage}}\par}} 

\def\keywords#1{{
        \centering{\begin{minipage}{4.5in}\footnotesize\baselineskip=10pt
        {\footnotesize\it Keywords}\/: #1
        \end{minipage}}\par}}

\newcommand{\bibit}{\nineit}
\newcommand{\bibbf}{\ninebf}
\renewenvironment{thebibliography}[1]
        {\frenchspacing
         \ninerm\baselineskip=11pt
         \begin{list}{\arabic{enumi}.}
        {\usecounter{enumi}\setlength{\parsep}{0pt}     
         \setlength{\leftmargin 12.7pt}{\rightmargin 0pt} 
         \setlength{\itemsep}{0pt} \settowidth
        {\labelwidth}{#1.}\sloppy}}{\end{list}}

\newcounter{itemlistc}
\newcounter{romanlistc}
\newcounter{alphlistc}
\newcounter{arabiclistc}

\newcommand{\fcaption}[1]{
        \refstepcounter{figure}
        \setbox\@tempboxa = \hbox{\footnotesize Fig.~\thefigure. #1}
        \ifdim \wd\@tempboxa > 5in
           {\begin{center}
        \parbox{5in}{\footnotesize\smalllineskip Fig.~\thefigure. #1}
            \end{center}}
        \else
             {\begin{center}
             {\footnotesize Fig.~\thefigure. #1}
              \end{center}}
        \fi}

\newcommand{\tcaption}[1]{
        \refstepcounter{table}
        \setbox\@tempboxa = \hbox{\footnotesize Table~\thetable. #1}
        \ifdim \wd\@tempboxa > 5in
           {\begin{center}
        \parbox{5in}{\footnotesize\smalllineskip Table~\thetable. #1}
            \end{center}}
        \else
             {\begin{center}
             {\footnotesize Table~\thetable. #1}
              \end{center}}
        \fi}

\def\@citex[#1]#2{\if@filesw\immediate\write\@auxout
        {\string\citation{#2}}\fi
\def\@citea{}\@cite{\@for\@citeb:=#2\do
        {\@citea\def\@citea{,}\@ifundefined
        {b@\@citeb}{{\bf ?}\@warning
        {Citation `\@citeb' on page \thepage \space undefined}}
        {\csname b@\@citeb\endcsname}}}{#1}}

\newif\if@cghi
\def\cite{\@cghitrue\@ifnextchar [{\@tempswatrue
        \@citex}{\@tempswafalse\@citex[]}}
\def\citelow{\@cghifalse\@ifnextchar [{\@tempswatrue
        \@citex}{\@tempswafalse\@citex[]}}
\def\@cite#1#2{{$\null^{#1}$\if@tempswa\typeout
        {IJCGA warning: optional citation argument 
        ignored: `#2'} \fi}}

\def\pmb#1{\setbox0=\hbox{#1}
        \kern-.025em\copy0\kern-\wd0
        \kern.05em\copy0\kern-\wd0
        \kern-.025em\raise.0433em\box0}

\def\fnt#1#2{\footnotetext{\kern-.3em
        {$^{\mbox{\scriptsize #1}}$}{#2}}}

\def\ps@myheadings{%
    \let\@oddfoot\@empty\let\@evenfoot\@empty
    \def\@evenhead{\slshape\leftmark\hfil}
    \def\@oddhead{\hfil{\slshape\rightmark}}
    \let\@mkboth\@gobbletwo
    \let\sectionmark\@gobble
    \let\subsectionmark\@gobble
    }
\font\tenrm=cmr10
\font\tenit=cmti10 
\font\tenbf=cmbx10
\font\bfit=cmbxti10 at 10pt
\font\ninerm=cmr9
\font\nineit=cmti9
\font\ninebf=cmbx9
\font\eightrm=cmr8

\def\qed{\hbox{${\vcenter{\vbox{                    
   \hrule height 0.4pt\hbox{\vrule width 0.4pt height 6pt
   \kern5pt\vrule width 0.4pt}\hrule height 0.4pt}}}$}}

\def\bsc{{\sc a\kern-6.4pt\sc a\kern-6.4pt\sc a}}       
\def\bflatex{\bf L\kern-.30em\raise.3ex\hbox{\bsc}\kern-.14em 
T\kern-.1667em\lower.7ex\hbox{E}\kern-.125em X} 

\begin{document}
\setlength{\textheight}{7.7truein}  

\thispagestyle{empty}

\markboth{\protect{\footnotesize\it Automatic Generation of Vacuum Amplitude}}{\protect{\footnotesize\it Automatic Generation of Vacuum Amplitude}}

\normalsize\textlineskip

\setcounter{page}{1}

\copyrightheading{}                     

\vspace*{0.88truein}

\centerline{\bf AUTOMATIC GENERATION OF VACUUM}
\vspace*{0.035truein}
\centerline{\bf AMPLITUDE MANY-BODY PERTURBATION SERIES}

\vspace*{0.37truein}
\centerline{\footnotesize P D STEVENSON}
\baselineskip=12pt
\centerline{\footnotesize\it Department of Physics, University of
  Surrey}
\baselineskip=10pt
\centerline{\footnotesize\it Guildford, Surrey, GU2 7XH, United Kingdom}
\centerline{\footnotesize\it E-mail: p.stevenson@surrey.ac.uk}


\vspace*{0.225truein}
\publisher{(received date)}{(revised date)}

\vspace*{0.25truein}
\abstracts{An algorithm and a computer program in Fortran 95 are
  presented which enumerate the Hugenholtz diagram representation of
  the many-body perturbation series for the ground state energy with a
  two-body interaction.  The output is in a form suitable for
  post-processing such as automatic code generation.  The results of a
  particular application, generation of \LaTeX\ code to draw the
  diagrams, is shown.}{}{} 

\vspace*{5pt}

\vspace*{1pt}\textlineskip      
\section{Many-Body Perturbation Theory}         
\vspace*{-0.5pt}
\noindent
Many problems of interest in quantum mechanics have no known analytic
solution, and some approximation is necessary to obtain a solution.  A
particularly useful approximation method is perturbation theory, which
consists of finding a way of separating the Hamiltonian into a part
which a solvable,  either numerically or analytically, and a part
which is small (See e.g. \cite{Mes} for a general treatment).  One may
then start from the solution of the solvable part and build in
successive orders of the perturbation in a power series approach to
obtain a closer approximation to the solution of the full problem,
providing the series converges.

In many-body quantum mechanics, the perturbation series is convenently
respresented in diagrammatic form, based on the work of Feynman\cite{Fey},
Goldstone\cite{Gol} and Hugenholtz\cite{Hug}.  Diagrams may be
prescriptively expressed in algebraic form which give a complete
representation of the perturbation series. One may construct such
series of diagrams and expressions for different observables in
Quantum Mechanics.  This paper is concerned 
with the ground state expectation value of the Hamiltonian operator
that is, the ground state energy, or vacuum amplitude for a
many-fermion system in the special, but common case that the
unperturbed problem is given by the solution of the  Hartree-Fock
equations and the system is governed by a two-body interaction.  This
situation is widely dealt with in textbooks \cite{Sza,Gro}. Each
order of correction in perturbation theory for the ground state energy
involves, in general, many diagrams and it is important to be able to
enumerate them, order by order, if one is to perform a
calculation\cite{Ste}.   

\section{Diagrammatic Representation}
The perturbation series for the ground state energy can be given by
diagrams in the Hugenholtz vein constructed according to the following
rules (see \cite{Sza} for a more detailed explanation):

For an $n^{th}$ order diagram, $n$ dots (vertices) are drawn in a
column.  These vertices are connected by directed lines subject to the conditions
\begin{enumerate}
\item Each vertex has two lines pointing in and two pointing out
\item Each diagram is connected, i.e. one must be able to go from any
  one vertex to any other by following some number of lines
\item No line connects a vertex with itself
\item Each diagram is topologically distinct
\end{enumerate}
For each diagram, a simple prescription exists to go from the pictorial
representation to an algebraic expression for the contribution to the
total energy from the particular diagram\cite{Sza}.

\begin{figure}[bht]
\centerline{\includegraphics{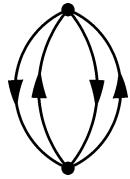}}
\caption{Second order diagram\label{fig:secondorder}}
\end{figure}

\begin{figure}[thb]
\centerline{\includegraphics{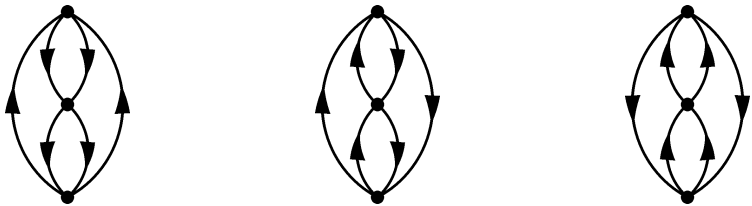}}
\caption{Third order diagrams\label{fig:thirdorder}}
\end{figure}

Following these rules, one can determine that there are no first order
diagrams since a line may not link a vertex with itself.  There is one
second order diagram, which is shown in figure \ref{fig:secondorder}
and three third order diagrams, which are shown in figure \ref{fig:thirdorder}.

\section{Diagram finding algorithm \label{sec:algorithm}}
A diagram of a given order $n_{ord}$ can be fully described by taking
all possible (unordered) pairs of distinct vertices, of which there
will be
\begin{equation}
n_{pairs} = \frac{1}{2}n_{ord}(n_{ord}-1)
\end{equation}
and specifying the number of lines linking each pair, along with the
direction of each line.   It is clear that the since the number of
lines connected to a given vertex must be four, that no more than four
lines may link any pair.  In fact, the case of four lines can exist
only in the second order diagram shown in figure \ref{fig:secondorder}
since any two vertices connected by 4 lines can not be connected to any
other vertices in the diagram, which is inadmissible by point 2 in the
above list of rules.  Given these facts, the following algorithm can
be used to find all possible diagrams:
\begin{itemize}
\item Create an ordered list of $n_{pairs}$ numbers with a
  specification for which vertices are associated with each pair: 1\&2,
  1\&3, $\ldots$, 1\&$n_{ord}$, 2\&3 , 2\&4 , $\ldots$, 2\&$n_{ord}$,
  $\ldots$, $(n_{ord}-1)$\&$n_{ord}$.
\item Allow each number in the list to take on independently the
  values 0,1,2,3 corresponding to the number of lines linking the
  pairs.  There will be $n_{pairs}^4$ such combinations.  The case of
  $n_{order}=2$ is treated specially by allowing the number of lines
  linking pairs to take the value 4.
\item For each combination of number of lines linking pairs,  rule 2
  from the above list is   checked along with the condition that four
  lines emanate from each vertex (due to rule 1).  Any combination not
  satisfying the rules is  rejected.  Any   combination which does
  satisfy these rules is a valid {\it unlabelled}  diagram.  The
  diagram is labelled in all possible ways by adding   arrows to the
  lines in all possible permutations of upward- and downward-going
  arrows and checking these permutations against rule 1.  Those which
  pass are valid (labelled) Hugenholtz diagrams.  The diagram is
  completely specified by the ordered list of numbers of lines
  connecting the pairs, and by a second list which gives the number of
  such lines which are pointing up; the rest must point down.
\end{itemize}
This provides an exhaustive search and is guaranteed to find all
diagrams, at the cost of potential slowness as the number of
permutations to check grows rapidly with order.  The two lists of
numbers given in the output may be post-processed to give, for
example, Fortran code which implements the algebraic form of the
diagrams.  Note that condition 3 is automatically satisfied by
the fact that in the specification of the pairs, we don't consider
pairing a vertex with itself.  Similary condition 4 is automatically
satisfied by the specification of the problem in terms of vertices
being ordered in a column in space.

\section{Description of Code} 
A Fortran 95 code to calculate valid diagrams using the approach
described in the previous section has been written and is presented in
Appendix A. The code is described in terms of the subprograms as
follows  

\subsection{\tt program hugenholz} 
$n_{ord}$ is read in, from which
  $n_{pairs}$ is calculated.  The pairs structure is set up so that
  the first pair links vertices one and two and so on as
  described in section 3.  A structure of $n_{pairs}$ numbers is
  initialized with all numbers zero and passed to subroutine {\tt new}
  to enumerate the diagrams.

\subsection{\tt subroutine new}
Given a diagram specification,
  a loop is made over the first number of pairs over possible allowed
  values $(0\ldots 3 \mathrm{\ or\ } 4)$.  The remainder of the diagram
  specification is then passed recursively to the subroutine {\tt new}
  for the further numbers to be looped over.  If we are at the last
  element of the specification, the diagram is checked for consistency
  using the function {\tt consistent} and then printed out prepended
  by `{\tt +}'.  Finally a call is made to {\tt label} to find all
  ways of labelling the lines with arrows.

\subsection{\tt function consistent}This function, which returns
  a {\tt logical}, takes a possible configuration of lines connecting
  pairs of vertices, and checks first that it satisfies the condition
  from rule 1 that each
  vertex has four lines connected to it.  Then the diagram is checked
  for being fully connected by starting at the bottom of the diagram,
  at vertex 1 and following lines until all vertices have been
  reached, in which case the diagram is consistent, or all lines have been
  followed without reaching some vertices, in which case it is
  inconsistent with rule 2 and we reject it.

\subsection{\tt subroutine label} A new array is set up which stores
  the number of upward-pointing lines for each pair of vertices.  This
  is initialized to zero and passed to {\tt newlabel}.

\subsection{\tt subroutine newlabel}
Like {\tt new}, this subroutine loops over all possible numbers of
upward-pointing lines that each pair may admit given the number of
total lines.  The looping is again recursive, and once the last pair
of lines is reached, the labelled diagram is tested with the function
{\tt testlab}.  If it passes the number of upward-pointing arrows is
printed prepended by `{\tt *}'.

\subsection{\tt function testlab}
This function tests for rule 2.  The number of lines entering and
leaving each vertex is calculated.  If, for every vertex, both these
numbers are two, then the diagram is valid.

\section{Sample output}
The output of the program for second order diagrams is
\begin{verbatim}
+ 4
* 2
\end{verbatim}
which says that there is one unlabelled diagram, which has one pair
linked by four lines.   There is furthermore only one way of
labelling it, with two of the lines pointing up, as shown in figure
\ref{fig:secondorder}.   The output for $n_{ord}=3$ is
\begin{verbatim}
+ 222
* 020
* 111
* 202
\end{verbatim}
which says that there is again only one unlabelled diagram, which has
three pairs, each linked by two lines.  This time there are three
possible ways of labelling the lines with arrows;  {\tt 020} in which
all lines point down except those linking vertices 1 and 3, {\tt 111}
in which each pair of vertices is connected by one upgoing and one
downgoing line, and {\tt 202} in which all lines point up except those
linking vertices 1 and 3.  These three possibilities are shown in
figure \ref{fig:thirdorder} from left to right in the order given
here.

\begin{figure}[h]
\includegraphics[width=4.5in]{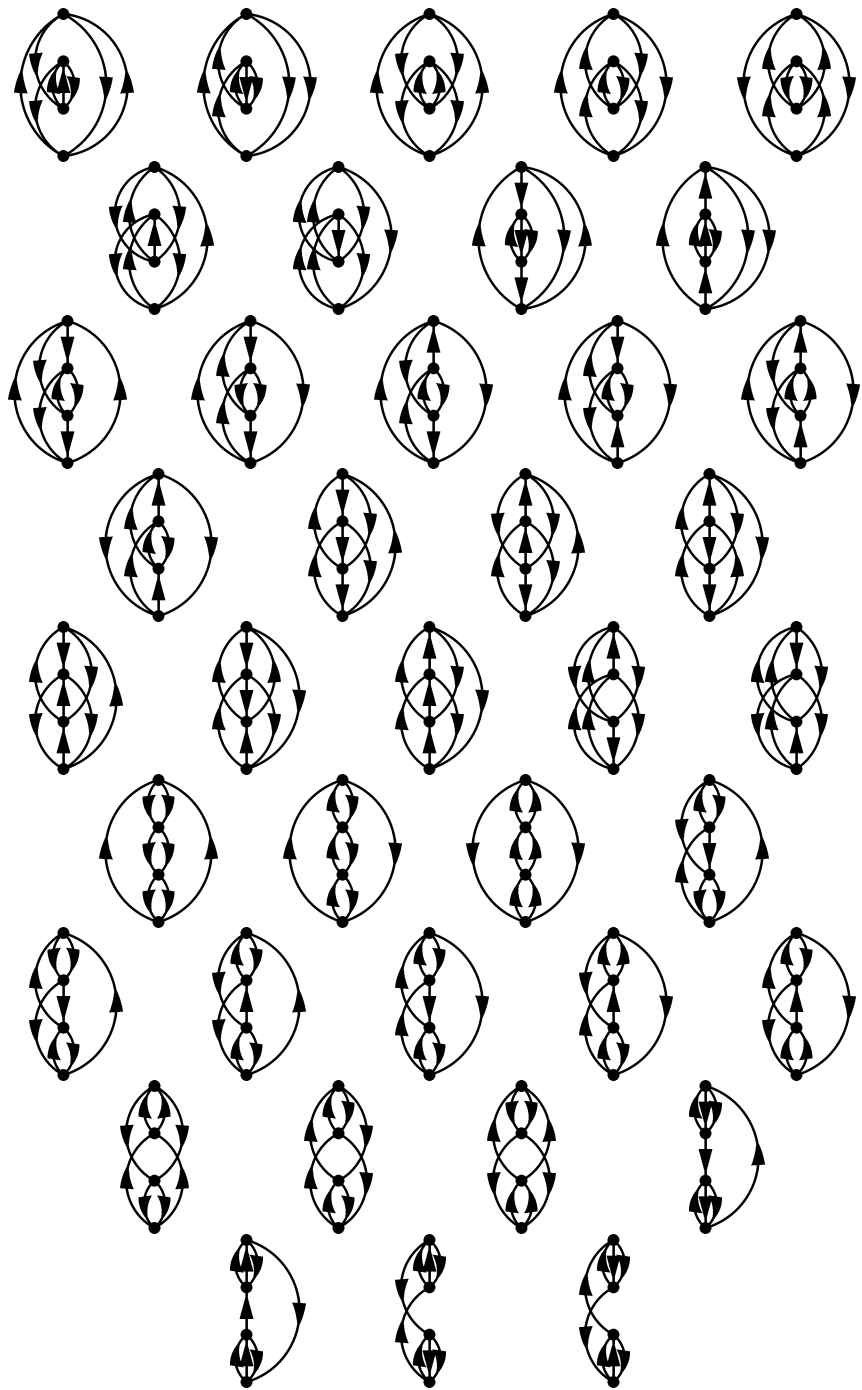}
\caption{The 39 fourth order diagrams\label{fig:fourthorder}}
\end{figure}

The graphical representation of the diagrams in this paper were
automatically generated from the output of the Fortran program {\tt
  hugenholtz} by means of a {\tt perl} program\footnote{available on
  request from the author}\,\, which outputs \LaTeX\ 
source using the FeynMF package\cite{Ohl}.  While it would have been
quite easy to work out second and third order diagrams by hand, the
ability to automatically enumerate and represent all diagrams for
higher orders becomes very useful.  For example, the number of
(labelled) diagrams at fourth, fifth, sixth and seventh orders are 39,
840, 27,300 and 1,232,280 respectively, which would be tedious to say
the least to enumerate by hand.  As an example of a less trivial
result, the fourth order diagrams, as generated from the output of
{\tt hugenholtz} are shown in figure \ref{fig:fourthorder}

\section{Summary}
An algorithm has been presented and implemented which enumerates and
represents Hugenholtz diagrams for the ground state perturbation series
for a many-body system interacting under the influence of a two-body 
interaction.  The representation is in a form suitable for
post-processing.  An example was given whereby \LaTeX code was
automatically generated to draw the diagrams.  A suggested further
application is automatic code generation to evaluate the series in
specific cases.

\nonumsection{Acknowledgements}
\noindent
This work was supported by the UK EPSRC

\appendix

\begin{verbatim}
module hug
  type pair
     integer :: v1,v2
  end type pair
  type diagram
     integer, dimension(:), pointer :: occ
  end type diagram
  integer :: npairs, norder

contains

  logical function consistent(diag,nlines,pairs)
    implicit none
    type(diagram), intent(in) :: diag
    integer, intent(in) :: nlines
    type(pair), dimension(:), intent(in) :: pairs
    integer :: linesum,  i, vertex, lines, j, start, fin
    integer, allocatable, dimension(:) :: link
    logical, allocatable, dimension(:) :: linked, to_try, tried
    logical :: isdiaglinked
  
    consistent = .true.
    ! check each vertex for correct number of lines:
    allocate(link(norder))
    link=0
    do i=1,size(diag%occ(:))
       link(pairs(i)%v1)=link(pairs(i)%v1)+diag%occ(i)
       link(pairs(i)%v2)=link(pairs(i)%v2)+diag%occ(i)
    end do
    do i=1,norder
       if(link(i)/=4) then
          consistent=.false. ; return
       end if
    end do
    deallocate(link)    
    ! Test for unlinked diagrams.    
    isdiaglinked=.false.
    allocate(linked(norder),to_try(norder),tried(norder))
    linked=.false.   ; to_try=.false.   ; tried=.false.
    linked(1)=.true. ; to_try(1)=.true.
    do 
       vertex=0
       do i=norder,1,-1
          if(to_try(i).eqv..true.) vertex=i
       end do
       if(vertex==0) exit
       do i=1,size(pairs)
          if(diag%occ(i)/=0) then
             if(pairs(i)%v1==vertex) then
                linked(pairs(i)%v2) = .true.
                if(tried(pairs(i)%v2) .eqv. .false.) then
                   to_try(pairs(i)%v2) = .true.
                end if
             else if(pairs(i)%v2==vertex) then
                linked(pairs(i)%v1) = .true.
                if(tried(pairs(i)%v1) .eqv. .false.) then
                   to_try(pairs(i)%v1) = .true.
                end if
             end if
          end if
       end do
       tried(vertex)=.true.
       to_try(vertex)=.false.      
       isdiaglinked=.true.
       do i=1,norder
          if(linked(i).eqv..false.) isdiaglinked=.false.
       end do
    end do
    deallocate(linked, to_try, tried)
    if(isdiaglinked.eqv..false.) consistent=.false.
  end function consistent

  subroutine label(diag, pairs)
    implicit none
    type(diagram) :: diag
    type(pair), dimension(:), intent(in) :: pairs
    integer, dimension(:), allocatable :: nups
    integer :: offset = 0
  
    allocate(nups(npairs))
    call newlabel(offset,nups,diag%occ(:), pairs)
    deallocate(nups) 
  end subroutine label

  logical function testlab(nuparray, pairs, occarray)
    implicit none
    type(pair), dimension(:), intent(in) :: pairs
    integer, dimension(:), intent(in) :: nuparray, occarray
    integer :: vstart, vend, i
    integer, dimension(:), allocatable :: enter

    allocate(enter(norder))
    enter=0
    do i=1,npairs
       vstart = pairs(i)%v1
       vend   = pairs(i)%v2
       if (vstart < vend) then
          enter(vend)   = enter(vend)+nuparray(i)
          enter(vstart) = enter(vstart)+ occarray(i)-nuparray(i)
       else
          enter(vstart) = enter(vstart)+nuparray(i)
          enter(vend)   = enter(vend)+ occarray(i)-nuparray(i)
       end if
    end do
    testlab=.false.
    if(all(enter==2)) testlab = .true.
    deallocate(enter)
  end function testlab

  recursive subroutine newlabel(offset, nuparray, occarray, pairs)
    implicit none
    integer :: offset
    integer, dimension(:) :: nuparray, occarray
    integer :: nuplo, nuphi, n, j
    type(pair), dimension(:), intent(in) :: pairs

    if(offset==npairs) then
       if (testlab(nuparray, pairs, occarray).eqv..true.) then
          write (*,'(a1,1x,100(i1))') '*',(nuparray(j),j=1,npairs) 
       end if
       return
    end if
    offset=offset+1
    nuplo = max(occarray(offset)-2,0)
    nuphi = min(occarray(offset),2)
    do n = nuplo,nuphi
       nuparray(offset) = n
       call newlabel(offset,nuparray,occarray, pairs)
    end do
    offset=offset-1
  end subroutine newlabel

recursive subroutine new(nlines, offset, temp, pairs) 
  implicit none
  integer, intent(in) :: nlines
  type(diagram) :: temp
  integer, intent(inout) :: offset
  type(pair), dimension(:), intent(in) :: pairs
  integer :: i, maxlinks = 3, j, ilo = 0

  if(offset==nlines/2-1 .and.sum(temp%occ(1:offset)).ne.4) return
  if(offset==npairs) then
     ! We have got to the end of the array - check for consitency:
     if(consistent(temp,nlines,pairs)) then
        write (*,'(a1,1x,100(i1))') '+',(temp%occ(j),j=1,npairs) 
        call label(temp, pairs)
     end if
     return
  end if
  if(npairs==1) maxlinks = 4
  do i=ilo,maxlinks
     temp%occ(1+offset)=i
     if(sum(temp%occ(1:offset+1))>nlines) cycle
     offset=offset+1
     call new(nlines, offset, temp, pairs)
     offset=offset-1
  end do
end subroutine new
end module hug

program hugenholtz
  use hug
  implicit none
  integer :: nfixed, nlines, index, i, j, offset = 0
  type(pair), dimension(:), allocatable :: pairs
  type(diagram) :: temp

  write(unit=*,fmt='("Input order of diagrams: ")',advance='no')
  read(unit=*,fmt=*) norder
  npairs = norder * (norder - 1) / 2
  allocate(pairs(1:npairs),temp%occ(1:npairs))
  nlines = 2 * norder
  index = 1
  do i = 1, norder
     do j = i + 1, norder
        pairs(index)%v1 = i ;  pairs(index)%v2 = j
        index = index + 1
     end do
  end do
  temp%occ(:) = 0
  call new(nlines, offset, temp, pairs)
  deallocate(temp%occ,pairs)
end program hugenholtz


\end{verbatim}

\nonumsection{References}


\begin{thebibliography}{000}
\bibitem{Mes}
A. Messiah, {\bibit Quantum Mechanics} (North Holland, Amsterdam, 1966)

\bibitem{Fey}
R. P. Feynman, {\bibit Phys. Rev. }{\bf 76}, 749 (1949), 769 (1949)

\bibitem{Gol}
J. Goldstone, {\bibit Proc. Roy. Soc. }{\bf A239}, 267 (1957)

\bibitem{Hug}
N. M. Hugenholtz, {\bibit Physica }{\bf 23}, 481 (1957)

\bibitem{Sza}
A. Szabo and N. Ostlund, {\bibit Modern Quantum Chemistry} (Dover, New
York, 1996)

\bibitem{Gro}
E. K. U. Gross, E. Runge and O. Heinonen, {\bibit Many-particle
  Theory} (Adam Hilger, Bristol, 1991)

\bibitem{Ste}
P. D. Stevenson, M. R. Strayer and J. Rikovska Stone, {\bibit
  Phys. Rev. }{\bf C63}, 054309 (2001)

\bibitem{Ohl}
T. Ohl, {\bibit Comput. Phys. Commun. }{\bibbf 90}, 340 (1995)

\end{thebibliography}
\end{document}